\documentclass{article} 
\title{Scalable Inference for Markov Processes with Intractable Likelihoods}
\author{Jamie Owen \and Darren J. Wilkinson \and Colin S. Gillespie}

\date{\vspace{-5ex}} 
\usepackage[margin=1.3in]{geometry}

\usepackage{algorithm}
\usepackage{amsmath}
\usepackage{graphicx}
\usepackage{natbib}
\usepackage{microtype}
\graphicspath{{fig/}}
\usepackage{color}
\usepackage{caption}
\usepackage{multirow}

\newcommand{\red}[1]{#1}

\usepackage{hyperref}
\begin{document}
\maketitle
\begin{center} 
{School of Mathematics \& Statistics, Newcastle University, \\ Newcastle upon Tyne, NE1 7RU, UK}
\end{center}

\begin{abstract}
\label{sec:abstract}

Bayesian inference for Markov processes has become increasingly relevant in
recent years. Problems of this type often have intractable likelihoods and prior
knowledge about model rate parameters is often poor. Markov Chain Monte Carlo
(MCMC) techniques can lead to exact inference in such models but in practice can
suffer performance issues including long burn-in periods and poor mixing. On the
other hand approximate Bayesian computation techniques can allow rapid
exploration of a large parameter space but yield only approximate posterior
distributions. Here we consider the combined use of approximate Bayesian
computation (ABC) and MCMC techniques for improved computational efficiency
while retaining exact inference on parallel hardware.


\end{abstract}

\section{Introduction} \label{sec:intro}

Stochastic kinetic models describe the probabilistic evolution of a dynamical
system. Motivated by the need to incorporate intrinsic stochasticity in the
underlying mechanics of the systems, stochastic models are increasingly used to
model a wide range of real-world problems, particularly, but not limited to,
computational systems biology, predator--prey population models, and single
species population development
\citep{kitano2002computational,boys2008bayesian,gillespie2010bayesian}. Systems
are governed by a network of reactions which change the state of the system by a
discrete amount, and hence are most naturally represented by a continuous time
Markov process on a discrete state space. Typically, exact realisations from the
model are obtained using the direct method \citep{gillespie1977exact}. There are
a number of faster, but approximate algorithms, such as the diffusion
approximation, moment closure, or hybrid simulation strategies
\citep{gillespie2000chemical,Gillespie2009a,Salis2005}.
 
Our goal is to perform statistical inference for the parameters that govern the
dynamics, where data are partially missing and prior knowledge of reaction rates
may be poor. Likelihood functions for such problems are rarely analytically
tractable but it is possible to leverage particle Markov chain Monte Carlo
(pMCMC) methods to perform exact inference
\citep{golightly2011bayesian,andrieu2010particle}. Typically, pMCMC algorithms
exhibit a high computational cost, since at each iteration we run a sequential
Monte Carlo (SMC) filter, which requires multiple forward simulations. This cost
can become overly prohibitive if convergence is slow. 

Approximate Bayesian computation (ABC) techniques, allow posterior inference to
be made for problems where evaluation of the likelihood function is unavailable,
relying heavily on the ability to simulate from the model
\citep{tavare1997inferring,pritchard1999population,beaumont2002approximate}. An
approximation to the true posterior distribution of interest is made via samples
of parameter vectors, $\theta$, that yield simulated data deemed ``close'' to the
observed data. Here the definition of close is controlled by a tolerance,
$\epsilon$, and distance function $\rho(\cdot)$, hence retained samples are from
$\pi(\theta|\rho(\mathcal{D},\mathcal{D}^{\ast}) \le \epsilon)$. The simple
rejection algorithm typically leads to high numbers of proposed parameter
vectors being rejected for meaningful tolerance values
\citep{pritchard1999population}. Further developments within this framework lead
to MCMC schemes which avoid calculation of the likelihood, as well as sequential
Monte Carlo algorithms, which typically exhibit better acceptance rates than the
simple rejection sampler
\citep{marjoram2003markov,del2006sequential,sisson2007sequential,toni2009approximate}.
Such techniques have successfully been applied to stochastic kinetic models for
approximate inference \citep{drovandi2011estimation,fearnhead2012constructing}.

In this article, we explore a scenario where we desire exact posterior inference
for model rate parameters, while exploiting parallel hardware. By coupling
scalable, approximate and exact techniques, we are able to increase
computational efficiency of posterior sampling via parallel particle MCMC and
improve convergence properties of chains that suffer expensive burn--in periods
to further increase efficiency.

\section{Stochastic Kinetic models} \label{sec:pmcmc}

Consider a network of reactions involving a set of $u$ species ${\cal X}_{1},
\ldots, {\cal X}_{u}$ and $v$ reactions $\mathbf{R}_{1}, \ldots, \mathbf{R}_{v}$
where each reaction $\mathbf{R}_{i}$ is given by
\begin{equation}  \label{eq:reaction}
  \mathbf{R}_{i}: \quad p_{i,1}{\cal X}_{1} + \ldots + p_{i,u}{\cal X}_{u} \rightarrow q_{i,1}{\cal X}_{1} + \ldots + q_{i,u}{\cal X}_{u}.
\end{equation}
Let $P$ be the $v \times u$ matrix of pre--reaction coefficients of reactants
$p_{i,j}$, similarly $Q$, the $v \times u$ matrix of post--reaction coefficients
of products. The stoichiometry matrix, defined as
\begin{equation} \label{eq:stoichiometry}
  S = (Q-P)^{\prime},
\end{equation}
is a useful way to encode the structure of the reaction network. Also define $X_{t}$ to
be the vector $(X_{1,t},\ldots,X_{u,t})$ denoting the number of species ${\cal
  X}$ present at time $t$.

Each reaction $\mathbf{R}_{i}$ is assumed to have an associated rate constant,
$\theta_{i}$, and a hazard function $h_{i}(X_{t},\theta_{i})$ which gives the
overall propensity for a type $i$ reaction to occur. The form of
$h_{i}(X_{t},\theta_{i})$ in many cases can be considered as arising from
interactions between components in a well mixed population. This leads to
reactions following the law of mass action kinetics where the hazard function
for a reaction of type $i$ takes the form
\begin{equation}
  \label{eq:mass}
  h_{i} = \theta_{i} \prod_{j=1}^{u} \left(
    \begin{array}{c}
      X_{j,t} \\
      p_{i,j}
    \end{array} \right).
\end{equation}
If $\theta = (\theta_{1}, \ldots, \theta_{v})$ and $h(X_{t},\theta) =
(h_{1}(X_{t},\theta),\ldots,$ \linebreak[4] $h_{v}(X_{t},\theta_{v}))$ then values for $\theta$
and $X_{0}$ complete full specification of this Markov process.

\subsection{The Direct method}\label{sec:gillespie}

Exact simulations of the evolution of such a system can be obtained via the
Direct method \citep{gillespie1977exact}. Algorithm~\ref{alg:gillespie} describes
the procedure for exact forward simulation of a stochastic kinetic model, given
its stoichiometry matrix, $S$, a vector of reaction rates, $\theta$, an initial
state vector $X_{0}$ and an associated hazard function $h(X_{t},\theta)$.
Reactions simulated to occur via the \red{Direct method} incorporate the
discrete and stochastic nature of the system since each reaction, which
increments or decrements species levels by discrete amounts, are chosen
probabilistically.

Less computationally expensive simulation algorithms such as the chemical
Langevin equation (CLE) relax the restriction
imposed by the discrete state space but retain the stochasticity of the
underlying mechanics, giving approximate realisations of the progression of the
species \citep{gillespie2000chemical}. For the purpose of exact inference and
this article we consider only exact \red{realisations of the reaction network via the direct method}. For a more in depth background into stochastic
kinetic modelling see \cite{wilkinson2011stochastic}.

\begin{algorithm}[t]
  \caption{The Direct method}
  \label{alg:gillespie}
  \begin{enumerate}
  \item Set $t=0$. Initialise the rate constants $\theta$ and initial states $X_{0}$.
  \item Calculate the hazard functions $h(X_{t},\theta)$ and $h_{0}(X_{t},\theta)
    = \sum_{i}^{v} h_{i}(X_{t},\theta_{i})$.
  \item Set $t = t + \delta t$ where
    \[
    \delta t \sim \operatorname{Exp}(h_{0}(X_{t},\theta)).
    \]
  \item Simulate the reaction index $j \in (1,\ldots,v)$ with probabilities $p_{j}$
    \[
    p_{j} = \frac{h_{j}(X_{t},\theta)}{h_{0}(X_{t},\theta)} \;.
    \]
  \item Set $X_{t+\delta t} = X_{t} + S[j]$ where $S[j]$ is the
    $j^\text{th}$ column of the stoichiometry matrix $S$.
  \item If $t < T$ return to 2.
  \end{enumerate}
  
\end{algorithm}

\red{
We anticipate data, $\mathcal{D} = (d_{0},d_{1},\ldots,d_{T})$, to be a collection of noisy, possibly partial observations of the system $X = (X_{0},X_{1},\ldots,X_{T})$ at discrete time intervals. 
}
\section{Methods for Bayesian Inference in Intractable Likelihood Problems}
\label{sec:methods}

Methods for likelihood--free Bayesian inference have become popular due to the
increasing use of models in which the likelihood function is either deemed too
computationally expensive to compute, or analytically unavailable. In
particular, approximate Bayesian computation (ABC), and particle filtering methods
have proven to be two classes of algorithms in which one can proceed with
inference in a scenario where evaluation of the likelihood is unavailable,
commonly termed likelihood--free inference.

\subsection{Approximate Bayesian computation} \label{sec:approximate}

The goal of approximate Bayesian computation techniques is to obtain a
collection of samples from the posterior distribution $\pi(\theta|\mathcal{D})$
where the likelihood function $\pi(\mathcal{D}|\theta)$ is unavailable, due to
either high computational cost or analytical intractability. Typically we rely
on the assumption that simulation from the model given a set of parameters,
$\theta$, is relatively straightforward, as in algorithm~\ref{alg:gillespie}.
Given a set of proposed parameter vectors we would ideally keep any such vectors
which yield data simulations that are equal to the observations that we have. In
reality however, we will typically match the simulated data to the observations
perfectly with low probability. To avoid rejecting all proposals, we instead
keep parameters that yield simulations deemed sufficiently close to the
observations. We characterise a set of simulated data, $\mathcal{D}^{\ast}$ as
sufficiently close if, for a given metric, $\rho(\cdot)$, the distance between
$\mathcal{D}^{\ast}$ and observed data $\mathcal{D}$ is below a tolerance
threshold, $\epsilon$. The simple rejection sampler is described in
algorithm~\ref{alg:abc}.

\begin{algorithm}[t]
  \caption{ABC rejection sampler} \label{alg:abc}
  \begin{enumerate}
  \item Generate a candidate parameter vector $\theta^{\ast} \sim \pi(\theta)$.
  \item Simulate a candidate data set $\mathcal{D}^{\ast} \sim
    \pi(\mathcal{D}|\theta^{\ast})$.
  \item Calculate a measure of distance between the candidate data,
    $\mathcal{D}^{\ast}$, and the observed data $\mathcal{D}$,
    $\rho(\mathcal{D},\mathcal{D}^{\ast})$.
  \item Accept $\theta^{\ast}$ if $\rho(\cdot) < \epsilon$ for some predetermined, fixed, $\epsilon$.
  \item Go to 1.
  \end{enumerate}
\end{algorithm}

Rather than leading to a sample from the exact posterior distribution, the
samples instead have the distribution
$\pi(\theta|\rho(\mathcal{D},\mathcal{D}^{\ast}) < \epsilon)$. As $\epsilon
\rightarrow 0$ this tends to the true posterior distribution if $\rho(\cdot)$ is
a properly defined metric on sufficient statistics. Conversely as $\epsilon
\rightarrow \infty$ one recovers the prior distribution on the parameters. Since
it is usually the case that sufficient statistics are not available in this type
of problem, further approximation can be made by choosing a set of summary statistics\red{, $S(\cdot)$, 
which} are not sufficient, but it is hoped describe the data well. \red{A set of summary statistics is often used over the observed data in problems in which the dimension of data is large. This is to combat the well--known curse of dimensionality that leads to intolerably poor acceptance rates in this type of scheme. In this case we replace step 3 of algorithm~\ref{alg:abc} with a step that first calculates $S(\mathcal{D}^{\ast})$ then assigns distance according to $\rho(S(\mathcal{D}),S(\mathcal{D}^{\ast}))$. The criteria for forming a good choice of a set of informative summary statistics is beyond the scope of this article and we refer the reader to the following paper which summarises the current research in this area, \cite{blum2013comparative}. Throughout this article we use a Euclidean metric on the simulated data points as we found that this gave sufficient acceptance rates.} If a large
value of $\epsilon$ is chosen, a large number of proposed candidates are
accepted, but little is learnt about the posterior distribution. Choosing a
small tolerance yields a much better approximation to $\pi(\theta|\mathcal{D})$
but at the expense of poorer acceptance rates and hence increased computational
cost. This rejection based approach has been included in a number of sampling
techniques.

Within the context of this article we consider a sequential approach to ABC
inference. An example of such a scheme, based on importance sampling, is described in
algorithm~\ref{alg:abcsmc}, see \cite{toni2009approximate} for details.

\begin{algorithm}[t]
  \caption{Sequential ABC} \label{alg:abcsmc}
  \begin{enumerate}
  \item Initialise $\epsilon_{0} > \epsilon_{1} > \ldots\ > \epsilon_{T} > 0$ and set the population indicator, $t=0$.
  \item Set particle indicator, $i=1$.
  \item If t = 0, sample $\theta^{\ast \ast} \sim \pi(\theta)$ \\*
    Else sample $\theta^{\ast}$ from the previous population $\{\theta^{(i)}_{t-1}\}$ with weights $w_{t-1}$ and perturb to obtain $\theta^{\ast \ast} \sim K_{t}(\theta|\theta^{\ast})$ \\*
    If $\pi(\theta^{\ast \ast}) = 0$, return to 3. \\*
    Simulate a candidate dataset $x^{\ast} \sim f(x|\theta^{\ast \ast})$ \\*
    If $d(x_{0},x^{\ast}) \ge \epsilon_{t}$, return to 3.
  \item Set $\theta_{t}^{(i)} = \theta^{\ast \ast}$ and calculate weight for particle $\theta_{t}^{(i)}$, $w_{t}^{(i)}$
    \[
    w_{t}^{(i)} = \left\{ 
      \begin{array}{cl}
        1, & \text{if } t = 0 \\
        \frac{\pi(\theta_{t}^{(i)})}{\sum^{N}_{j=1} w_{t-1}^{(j)} K_{t}(\theta^{(j)}_{t-1},\theta^{(i)}_{t})}, & \text{if } t > 0
      \end{array} 
    \right\}.
    \]
    If $i < N$ set $i = i+1$, go to 3
  \item Normalise the weights, if $t<T$, set $t=t+1$ and go to 2.
  \end{enumerate}
\end{algorithm}

Algorithm~\ref{alg:abcsmc} as described here allows us choice of an essentially
arbitrary perturbation kernel $K_{t}(.)$ for preventing degeneracy of particles,
and the sequence of tolerances $\epsilon_{0},\ldots,\epsilon_{T}$. In practice
the choice of $K_{t}(.)$ will have an important effect on the overall efficiency
of the algorithm. A perturbation which gives small moves will typically allow
acceptance rates to remain higher, whilst exploring the space slowly. Conversely
making large moves allows better exploration of the parameter space, but usually
at the cost of reduced acceptance rates. We shall use the `optimal' choice of a
random walk perturbation kernel for this algorithm as detailed in
\cite{filippi2013optimality}. For a multivariate Gaussian random walk kernel at
population $t$, we choose $\Sigma^{(t)}$ as
\begin{equation}
  \label{eq:optimal}
  \Sigma^{(t)} \approx \sum_{i=1}^{N} \sum_{k=1}^{N_{0}} \omega^{(i,t-1)} \tilde{\omega}^{(k)}(\tilde{\theta}^{(k)} - \theta^{(i,t-1)})(\tilde{\theta}^{(k)} - \theta^{(i,t-1)})^{\prime}
\end{equation}
where $N$ is the number of weighted samples we have from the previous population
$\pi(\theta|\rho(\mathcal{D},\mathcal{D}^{\ast}) < \epsilon_{t-1})$ with weights
$\omega$, $N_{0}$ is the number of those samples, $\tilde{\theta}$, that satisfy
$\pi(\theta|\rho(\mathcal{D},\mathcal{D}^{\ast}) < \epsilon_{t})$ with
associated normalised weights $\tilde{\omega}$. This builds on the work of
\cite{beaumont2009adaptive}.

The choice of sequence of tolerances $\epsilon_{0},\ldots,\epsilon_{T}$ also has
a large effect on the efficiency of the algorithm. A sequence which decreases
too slowly will yield high acceptance rates but convergence to the posterior
will be slow. Whereas a rapidly decreasing sequence typically yields poor
acceptance rates. We will, throughout this article, use an adaptive choice of
$\epsilon_{t}$ determined by taking a quantile of the distances at $(t-1)$. \red{It should be noted that \cite{silk2013optimizing} observe that there can be issues with convergence when using an adaptive tolerance based on quantiles. However the authors are unaware of a generally accepted solution to this problem.}

\subsubsection{ABC methods in parallel}\label{sec:appeal}

Non MCMC--based ABC techniques, such as those described in
algorithm~\ref{alg:abc} and algorithm~\ref{alg:abcsmc} are often amenable to
parallelisation. Since all proposed parameters are independent samples from the
same distribution, and the acceptance step for each does not depend on the
previously accepted particle, the bulk of computation for a sequential ABC
scheme can be run in parallel, greatly reducing overall CPU time required to
obtain a sample from the target. Some thread communication is necessary as we
move from one distribution in the sequence to the next, something that would be
avoided if using the simple rejection sampler, but typically this is small in
comparison to the work done in forward simulation.

Coding of a parallel ABC algorithm adds little complexity over a
non-parallelised version. This is particularly true for the simple rejection
sampler where one effectively just runs a sampler on each of N processors and no
communication between threads is needed.

\subsection{Markov chain Monte Carlo}
\label{sec:markov}

Suppose interest lies in $\pi(\theta|\mathcal{D})$, and that we wish to
construct a MCMC algorithm whose stationary distribution is exactly this
posterior. Using an appropriate proposal function $q(\theta^{\ast}|\theta)$ we
can construct a Metropolis Hastings algorithm to do this as described in
algorithm~\ref{alg:mh}. This can often be impractical due to unavailability of
the likelihood term, $\pi(\mathcal{D}|\theta)$.

\begin{algorithm}[t]
  \caption{Metropolis Hastings} \label{alg:mh}
  \begin{enumerate}
  \item Initialise with a random starting value $\theta \sim \pi(\theta)$.
  \item Propose a move to a new candidate $\theta^{\ast} \sim q(\theta^{\ast}|\theta)$.
  \item Accept the move with probability
    \begin{equation}
      \min\left\{1, \frac{\pi(\mathcal{D}|\theta^{\ast})\pi(\theta^{\ast})q(\theta|\theta^{\ast})}
        {\pi(\mathcal{D}|\theta)\pi(\theta)q(\theta^{\ast}|\theta)} \right\},
    \end{equation}
    else remain at $\theta$.
  \item Return to 2.
  \end{enumerate}
\end{algorithm}

In practice one normally chooses a proposal function $q(\theta^\ast|\theta)$
such that proposed moves from $\theta$ to $\theta^\ast$ are distributed
symmetrically about $\theta$ giving rise to random walk Metropolis Hastings.
Typical choices include a uniform $\mathcal{U}(\theta - \sigma, \theta +
\sigma)$ or Gaussian $\mathcal{N}(\theta,\Sigma)$ distribution. In this article
we shall consider Gaussian innovations for random walk proposals. It has been
shown that under various assumptions about the target, the optimal scaling for a
random walk Metropolis algorithm using a Gaussian proposal, $\Sigma_q$, for a
$d$ dimensional distribution is
\begin{equation}
  \Sigma_{q} = \frac{(2.38)^2}{d}\Sigma,
\end{equation}
where $\Sigma$ is the posterior covariance \citep{roberts1997weak,roberts2001optimal}. Typically, however, the posterior covariance is unavailable.

\subsubsection{Parallel MCMC}
\label{sec:parallel}

MCMC algorithms are somewhat less amenable to parallelisation than rejection
sampling algorithms. Essentially there are two options to be considered,
parallelisation of a single MCMC chain or the construction of parallel
chains\red{. For} an in depth discussion see \cite{wilkinson2006parallel}. Parallelisation of a single chain in many
scenarios is somewhat difficult due to the inherently iterative nature of a
Markov chain sampler and will not be discussed in detail here. Running parallel
chains, however, is straightforward.

\subsubsection{Parallel chains}
\label{sec:parallechains}

In practice, chains initialised with an arbitrary starting point will be, after
an appropriate burn--in period, essentially sampling from the target. In the
context of a serial implementation there is still some debate over whether it is
better to run a single long chain, or many shorter chains. The benefits of the
single chain are that any burn in period is only suffered once, \red{whereas}
the argument for multiple shorter chains is that one may better diagnose
convergence to the stationary distribution. The argument changes when
considering parallel implementation. Indeed, if burn--in periods are relatively
short, running independent chains on separate processors can be a very time
efficient way of learning about the distribution of interest.

Burn--in is still a potential limiter on the scaling of the performance to be
had when employing parallel chains with the number of processors. The greater
the burn--in period of a chain, the more time each processor has to waste
computing samples that will eventually be thrown away. The theoretical speed up
calculation given $N$ processors to obtain $n$ stored samples with a burn--in
period $b$ is
\begin{equation}  \label{eq:speedup}
  \text{Speed-up}(N) = \frac{b+n}{b+\frac{n}{N}},
\end{equation}
which is clearly limited for any $b>0$, as $N \rightarrow \infty$
\citep{wilkinson2006parallel}. A ``perfect'' parallelisation of multiple chains
then is one in which we have no burn--in period, i.e initialise each chain with
an independent draw from the target. The performance gain when run on $N$
processors is then of factor $N$. Clearly this ``perfect'' situation will not be
possible to implement in practice since we typically are unable to sample from
the target.

\subsubsection{Particle MCMC} \label{sec:particle}

Re-consider the Metropolis Hastings MCMC algorithm from
section~\ref{sec:markov}. The ability to construct such a scheme relies on the
evaluation of the likelihood function $\pi(\mathcal{D}|\theta)$ among others.
Crucially, the problems of interest discussed in this article are such that the
likelihood term $\pi(\mathcal{D}|\theta)$ is unavailable. However
\cite{andrieu2009pseudo} proposed a pseudo marginal MCMC approach to this
problem. In the acceptance ratio of the MCMC scheme, replace the intractable
likelihood, $\pi(\mathcal{D}|\theta)$, \red{with a non--negative} Monte Carlo estimate. It can be
shown that, provided $E[\hat{\pi}(\mathcal{D}|\theta)] =
\pi(\mathcal{D}|\theta)$, the stationary distribution of the resulting Markov
chain is exactly the distribution of interest. In the context of inference for
state space models, it is natural to make use of sequential Monte Carlo
techniques through a bootstrap particle filter, \citep{doucet2001sequential}, to
obtain estimates of the likelihood $\hat{\pi}(\mathcal{D}|\theta)$. The
bootstrap particle filter for application to Markov processes is described in
algorithm~\ref{alg:boot}.

\begin{algorithm}[t]
  \caption{The boot--strap particle filter} \label{alg:boot}
  \red{Let $\mathbf{X}_{t}^{\ast}$ denote a set of $N$ particles, $\{(x_{t}^{i}, \pi_{t}^{i}) : i
  = 1, \ldots, N\}$. The filter assumes fixed parameters and so we drop the
  $\theta$ from our notation.
  \begin{enumerate}
  \item Initialise at $t=0$ and draw a set of $N$ independent samples,  $\mathbf{X}_{t}^{\ast} \sim \pi(X_{0})$.
  \item Sample a set of indices for candidates for forward simulation, $I_{t}^{i}$ according to the weights $\pi_{t}$.
  \item Simulate forward from the model the chosen paths, $x_{t+1}^{i} \sim \pi(x_{t+1}^{i}|x_{t}^{I^{i}_{t}})$.
  \item Calculate weights, $w_{t+1}^{i} = p(d_{t+1}|x_{t+1}^{i})$, and normalise to set $\pi_{t+1}^{i} = \frac{w_{t+1}^{i}}{\sum_{j=1}^{N} w_{t+1}^{j}}$.
  \item Set $t:= t + 1.$
  \item  If $t < T$, sample a new set of particles $\mathbf{X}_{t}^{\ast} \sim \pi(\mathbf{X}_{t}|D_{0:t})$ and return to 2.
  \end{enumerate}
  Define $\hat{p}(d_{t}|D_{1:t-1}) = \frac{1}{N}\sum_{i = 1}^{N} w_{t}^{i}$, then $\hat{p}(D_{1:t}) = \prod_{t=1}^{T} \hat{p}(d_{t}|D_{1:t-1})$.}
\end{algorithm}

Substituting the estimate of $\hat{\pi}(\mathcal{D}|\theta)$ in place of the
likelihood function yields a MH algorithm with acceptance probability
\begin{equation} \label{eq:pseudoacceptance}
  \min\left\{1, \frac{\hat{\pi}(\mathcal{D}|\theta^{\ast})\pi(\theta^{\ast})q(\theta|\theta^{\ast})}
        {\hat{\pi}(\mathcal{D}|\theta)\pi(\theta)q(\theta^{\ast}|\theta)} \right\}.
\end{equation}
It turns out that the particle filter's estimate of marginal likelihood is
unbiased, giving rise to an ``exact approximate" pseudo-marginal MCMC algorithm.
This use of a SMC sampler within a pseudo-marginal algorithm is a special case
of the particle marginal Metropolis Hastings (PMMH) algorithm described in
\cite{andrieu2010particle}, which in general can be used to target the full
joint posterior on the state and rate $\pi(\theta,X|\mathcal{D})$. In this
article we consider only the pseudo-marginal special case. As discussed in
section~\ref{sec:parallel} this type of scheme does not lend itself well to
parallelisation unless we have the opportunity to initialise chains with a
sample from the posterior distribution $\pi(\theta|\mathcal{D})$. We recognise
that rather sophisticated parallel particle filter algorithms which have the
potential for increasing the speed of moves within a single chain are available.
However the speed up is limited by necessary communication and synchronisation
between processors and typically do not scale well.

The pseudo--marginal MCMC scheme requires specification of a number of
particles, $N$, to be used in the boot--strap filter. It was noted, \citep{andrieu2009pseudo}, that the efficiency of the scheme decreases as the variance of the
estimated marginal likelihood increases. This can be overcome by increasing the
value of $N$, albeit at an increased computational cost. An optimal choice of
$N$ was the subject of \cite{pitt2012some} and \cite{doucet2012efficient}. The
former suggest that $N$ should be chosen such that the variance in the estimated
log--likelihood is around 1 while the latter show that the \red{efficiency penalty} is small for
values between 0.25 and 2.25.
    
\subsection{Combined use of ABC and pMCMC} \label{sec:combine}

Since it is typically not possible to initialise a MCMC chain with a draw from the desired target, we propose an approach to parallel MCMC by choosing initial parameter vectors according to samples from an approximate posterior distribution. The intuition is that if we have a reasonable approximation to the target of interest, samples from the approximation will closely match those from $\pi(\theta|\mathcal{D})$. Because of this we expect that after a very short burn--in we are sampling from the desired target. Hence we are approaching the scenario of \red{near perfect} parallelisation of MCMC. Clearly the better the approximation the shorter the burn--in for each chain will be.

If we first run an ABC scheme targeting an approximation to this distribution, we can exploit parallel hardware to eliminate a large region of prior parameter space very quickly. Then take a set of $N$ independent samples from $\pi(\theta|\rho(\mathcal{D},\mathcal{D}^{\ast}) < \epsilon)$ to initialise $N$ independent, parallel, pMCMC chains each targeting the exact posterior distribution $\pi(\theta|\mathcal{D})$.

We can in some sense consider this process of obtaining a sample from an ABC
posterior as an artificial burn--in period. Crucially however, since the ABC
algorithm yields a large set of samples from
$\pi(\theta|\rho(\mathcal{D},\mathcal{D}^{\ast}) \le \epsilon)$ the
computational price of performing the artificial burn--in has only to be paid
once and can be parallelised. With a set of samples we can initialise an arbitrary number of MCMC chains
each with an independent parameter vector $\theta_{0}$ which comes from a
distribution close to the target $\pi(\theta|\mathcal{D})$.

Since ABC is being used only for initialisation, it is not used as a prior \red{for the 
MCMC} algorithm, nor does it form part of the proposal. Hence, the ABC
approximation does not affect the exactness of the pMCMC target.

\subsection{Random walk pMCMC using ABC} \label{ssec:tune}

We now refer back to the optimal choice of Gaussian random walk kernel of
\cite{roberts2001optimal} mentioned in section~\ref{sec:markov}. Since we hope
that we start with a reasonable approximation to the true posterior, we likewise
consider that the covariance of a sample from
$\pi(\theta|\rho(\mathcal{D},\mathcal{D}^{\ast}) < \epsilon)$ denoted here
$\Sigma_{\texttt{ABC}}$, will be close to the covariance of true posterior
samples. This allows us to use the approximate posterior covariance structure as
an informative tool for calibrating the random walk innovations. In most MCMC
applications, since the true posterior variance is unknown, $\Sigma_{q}$ requires
tuning, typically from pilot runs. In the case of using $\Sigma_{\texttt{ABC}}$
we hope that we can also remove the necessity to tune $\Sigma_{q}$. In practice
we take
\begin{equation}
   \Sigma_{q} = \frac{(2.38)^2}{d}\Sigma_{\texttt{ABC}}.
\end{equation}
In addition we use our approximation to automatically tune the number of
particles required for the particle filter. In practice we take the average of
the approximate distribution, $\bar{\theta}_{\texttt{ABC}}$ to calculate the
number of particles required to satisfy
\begin{equation}
  \label{eq:particles}
  \operatorname{Var}(\log(\hat{\pi}(\mathcal{D}|\bar{\theta}_{\texttt{ABC}}))) \simeq 2,
\end{equation}
in line with \cite{sherlock2013efficiency}. \red{We do this by running the particle filter a number of times with varying numbers of particles until the condition is satisfied.}
The hybrid ABC pMCMC algorithm is outlined in algorithm~\ref{alg:hybrid}.

\begin{algorithm}[t]
  \caption{Hybrid ABC pMCMC}
  \label{alg:hybrid}
  \begin{enumerate}
  \item  Run an ABC algorithm targeting $\pi(\theta|\rho(\mathcal{D},\mathcal{D}^{\ast}) \le \epsilon_{T})$.
  \item Initialise multiple MCMC chain with a sample $\theta_{0} \sim
    \pi(\theta|\rho(\mathcal{D},\mathcal{D}^{\ast}) \le \epsilon_{T})$ and set
    $i = 1.$
  \item Propose $\theta^{\ast} \sim q(\theta^{\ast}|\theta)$ for $q(\cdot)$ a
    Gaussian random walk kernel, $\mu_{q} = \theta, \quad \Sigma_{q} =
    \frac{(2.38)^2}{d}\Sigma_{\texttt{ABC}}.$
  \item Approximate $\pi(\mathcal{D}|\theta^{\ast})$ via a bootstrap particle
    filter, $\hat{\pi}(\mathcal{D}|\theta^{\ast})$ where the number of particles
    is chosen as in (\ref{eq:particles}).
  \item With probability
    \[
    \min\left\{1,
      \frac{\hat{\pi}(\mathcal{D}|\theta^{\ast})\pi(\theta^{\ast})q(\theta|\theta^{\ast})}
      {\hat{\pi}(\mathcal{D}|\theta)\pi(\theta)q(\theta^{\ast}|\theta)} \right\}
    \]
    set $\theta^{(i)} = \theta^{\ast}$ else set $\theta^{(i)} = \theta^{(i-1)}$.
    \item Set $i:= i + 1$ and return to 3.
  \end{enumerate}
\end{algorithm}

\section{Applications} \label{sec:applications}

\subsection{The Lotka--Volterra system} \label{sec:lvintro}

The Lotka--Volterra predator--prey system is a simple stochastic kinetic model
\citep{lotka1925elements,volterra1926fluctuations}. The system is characterised
by a set of three reactions detailing interactions between two species; prey
$X$ and predators $Y$:
\begin{align}
  \centering
  \begin{array}{rrcl}
    \mathbf{R}_{1}: & X & \rightarrow & 2X \\
    \mathbf{R}_{2}: & X + Y & \rightarrow & 2Y \\
    \mathbf{R}_{3}: & Y & \rightarrow & \emptyset.
  \end{array}
  \label{eq:lvmodel}
\end{align}
The reactions $\mathbf{R}_{1}$, $\mathbf{R}_{2}$ and $\mathbf{R}_{3}$ can be thought of as prey birth, an interaction resulting in a prey death and predator birth, and predator death respectively. Whilst this is a relatively simple example of a model in this context, it highlights many of the difficulties associated with more complex systems. We summarise the system by its stoichiometry matrix, $S$ and hazard function $h(\mathbf{X}_{t},\theta)$:
\begin{align}
  S &= \left(
    \begin{array}{rrr}
      1 & -1 & 0 \\
      0 & 1 & -1
    \end{array} \right), &
  h(\mathbf{X}_{t},\theta) &= (\theta_{1} X_{t }, \theta_{2} X_{t} Y_{t}, \theta_{3}Y_{t}).
\end{align}
Figure~\ref{fig:lvproblems}(a) is a set of synthetic, noisy observations of the
two species simulated using the Gillespie algorithm with true parameter values
$\log(\theta) = (0,-5.30,-0.51)$ together with a Gaussian measurement error
structure, $d_{j,t} \sim \mathcal{N}(X_{j,t},10^{2})$.

\subsubsection{\red{pMCMC for Lotka--Volterra}}
\label{sec:pmcmc}

\begin{figure*}[t]
  \centering
  \includegraphics[width = \textwidth]{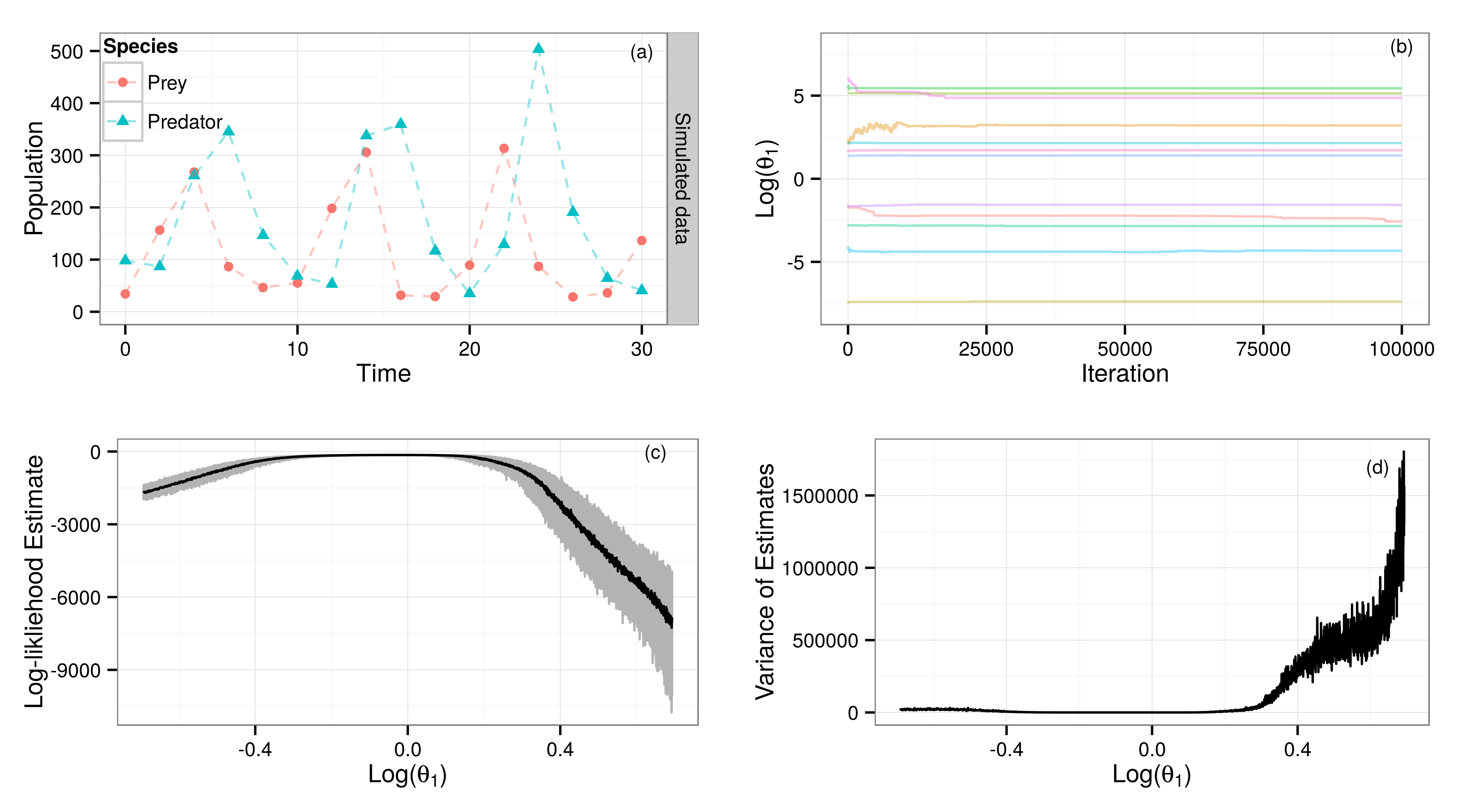}
  \caption{Investigating computational issues with pMCMC for the Lotka--Volterra
    model \red{defined in section~\ref{sec:lvintro}}. (a) The true underlying synthetic data set. Species are observed at
    discrete time points and corrupted with $N(0, 10^2)$ noise. (b) Twelve trace
    plots of $\theta_{1}$ from pMCMC chains initialised with random draws from
    the prior (see expression \ref{eq:prior}). The chains fail to explore the space. (c) shows the median and 95\% interval for estimates of the log--likelihood from the particle filter for varying $\theta_{1}$ close to the true value, for $\theta_{2}$ and $\theta_{3}$ fixed at the true values. (d) shows that the variance og log--likelihood estimates increases away from the true values.}
  \label{fig:lvproblems}
\end{figure*}

\begin{figure*}[t]
  \centering
  \includegraphics[width = \textwidth]{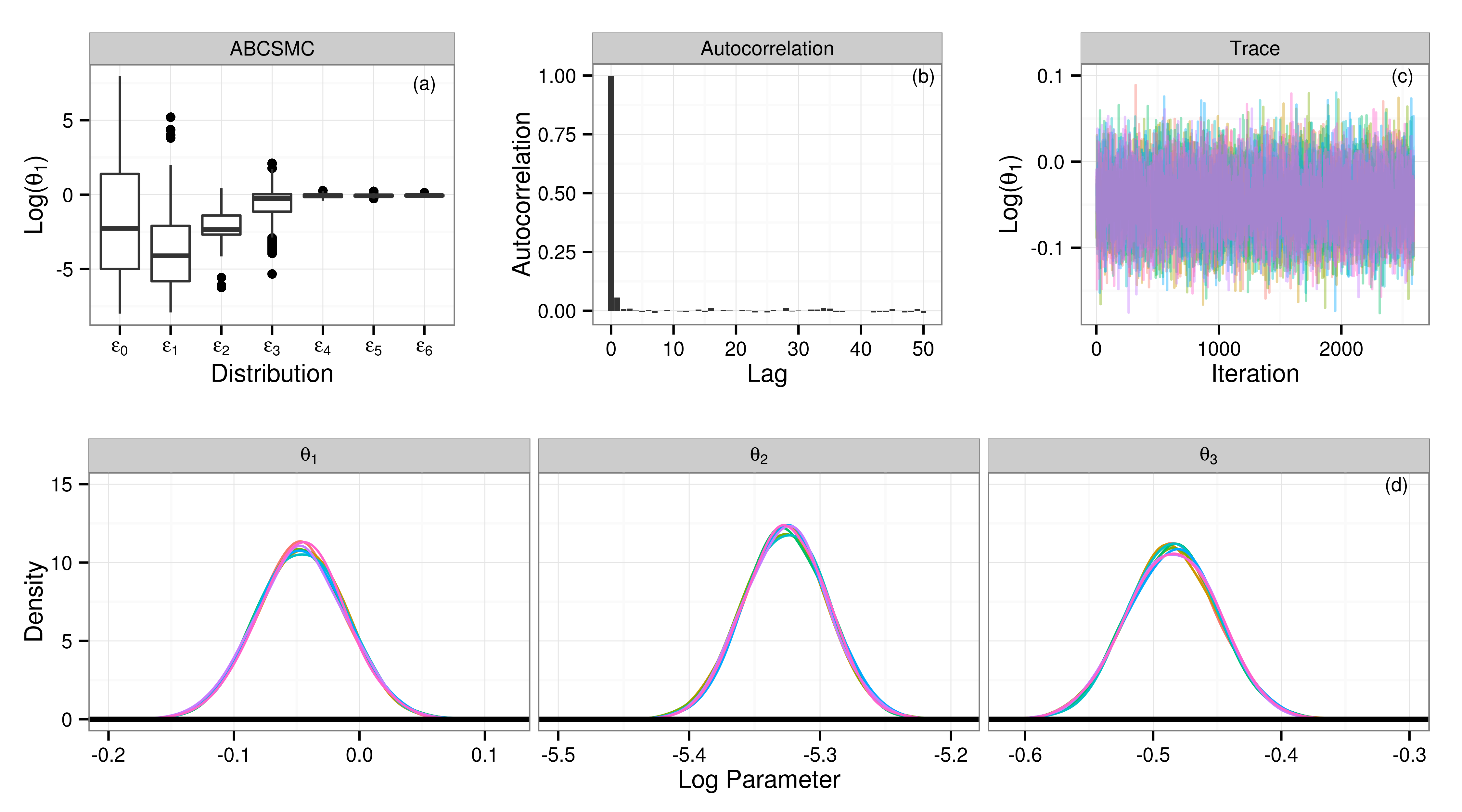}
  \caption{Analysis of results for the synthetic data for the Lotka--Volterra
    model. (a) are successive distributions of $\log(\theta_{1})$ in the
    sequential ABC scheme, algorithm~\ref{alg:abcsmc}. (b) show autocorrelations for chain 1, representative of each of the parallel chains, and (c) are traces
    of eight parallel MCMC chains for $\log(\theta_{1})$. Note that each chain
    is sampling from the same stationary distribution and mixing appears good.
    (d) are the posterior densities for $\log(\theta)$, each chain leads to a
    posterior density plot that is very close to that of every other chain. True
    values $\log(\theta = (0.0,-5.30,-0.51)$ are well identified}
  \label{fig:lvresults}
\end{figure*}

Exact parameter inference is possible for this system using \red{ a pMCMC} scheme; provided the chain is initialised near the posterior mode
\citep{wilkinson2011stochastic}. However under poor initialisation, a \red{pMCMC} scheme will perform very badly. 

Suppose we have little prior knowledge on the rate parameters,
\begin{equation}  \label{eq:prior}
  \log(\theta_{i}) \sim \mathcal{U}(-8,8), \qquad i = 1,2,3.
\end{equation}
We take a prior distribution for the initial state $X_{0}$ on each individual species as Poisson distributions with rate parameter equal to the true initial conditions,
\begin{equation}
  x_{1,0} \sim \operatorname{Pois}(50), \quad x_{2,0} \sim \operatorname{Pois}(100).
\end{equation}
Further we assume that the variance, $\sigma^{2}$, of the Gaussian measurement error is known.

Using a Gaussian random walk \red{proposal on $\log(\theta)$, $q(\log(\theta)^{\ast}|\log(\theta))$} we construct a
MH algorithm using a bootstrap particle filter, targeting
$\pi(\theta|\mathcal{D})$ as in algorithm~\ref{alg:mh}.
\red{Figure~\ref{fig:lvproblems}(b) shows that when initialising the chain using
random draws from the weakly informative prior, the chains do not explore the space, failing to converge. Further
investigation shows why this is happening. Figure~\ref{fig:lvproblems}(c and d) show that away from the true values, the variance of the log--likelihood estimates from the bootstrap particle filter increases sharply. Figure~\ref{fig:lvproblems} (c) shows the $95\%$ interval of log--likelihood estimates over 100 runs of a particle filter using 150 particles. Figure~\ref{fig:lvproblems} (d) shows that even close to the true value, $\theta_{1} = 1$, the variance of the log-likelihood estimates increases dramatically with a small move. In both figure~\ref{fig:lvproblems}(c) and (d) the other two parameters were kept fixed at the true values. When we are in a region with neligible likelihood the chain has a tendency to stick as a result of the variability in the likelihood estimates. }
On
top of this, small proposed moves can lead to large variation in the estimated
likelihood, leading to poor exploration of the tails in the posterior
distribution. Whilst we are guaranteed to eventually converge to the stationary
distribution, the required computational cost, without carefully thought out
initialisation, could be very high. We note that this is not a failure of the
theory or algorithm, \red{ but a consequence of the sensitivity to initialisation of parameter values experienced in this type of model}.

We therefore apply the proposed ABC initialisation for a ``perfect'' parallel pMCMC scheme.

\subsubsection{\red{Results for the Lotka--Volterra model using a hybrid approach}} \label{sec:resultslv}

We use a sequence of seven distributions
$\pi(\theta|\rho(\mathcal{D},\mathcal{D}^{\ast}) < \epsilon_{t}),\, t =
0,\ldots,6$ to obtain our approximation to the posterior. For each population
$t$ we take $\epsilon_{t}$ as the 0.3 quantile of the distribution of distances
from the samples at $t-1$ and propose candidates until we reach a sample of 1000
for each $t$. At each stage we perform the forward simulation of the model given
proposed parameter values in parallel. Figure~\ref{fig:lvresults} (a) shows
summaries for log of the reaction 1
rate parameter $\log(\theta_{1})$ for each of the distributions in the series through the sequential ABC. The
distributions quickly remove \red{a large region of space that, had we sampled from the prior distribution to initialise the chain, are likely to have been poor starting points. The scheme converges} around
the true value $\log(\theta_{1}) = 0$. Given the sample from
$\pi(\theta|\rho(\mathcal{D},\mathcal{D}^{\ast}) < \epsilon_{6})$ we initialise
eight MCMC chains on separate processors with random draws. The results in figure~\ref{fig:lvresults}(b,c,d) are
then, the 20,000 pooled samples from the eight independent parallel chains, each of which has
been thinned by a factor of 100 to give 2,500 samples. Each chain is sampling
from the same stationary distribution as seen in the trace plot for
$\theta_{1}$, figure~\ref{fig:lvresults} (c), and mixing is good,
figure~\ref{fig:lvresults} (b). Further the true parameter values, $\log(\theta)
= (0,-5.30,-0.51)$, used to simulate the data are well identified within the
posterior densities, figure~\ref{fig:lvresults} (d).

\subsection{Real-data problem -- aphid model}

\begin{figure*}[t]
  \includegraphics[width=\textwidth]{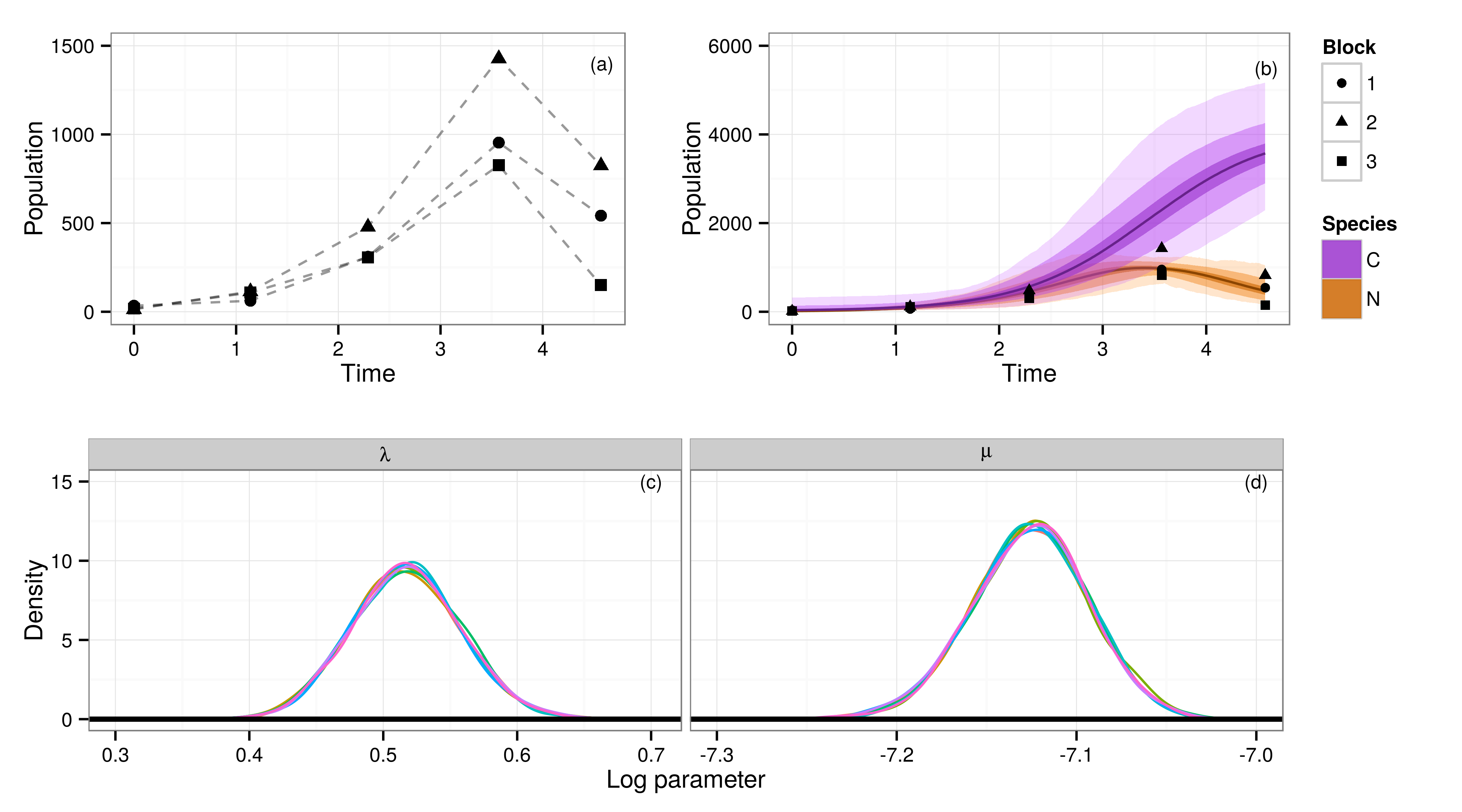}
  \caption{Analysis of the real data for the aphid growth model. (a) There are
    the three data sets of aphid counts each with of five observations. (b) The
    posterior predictive model fit given a sample from the collection of
    posterior densities. (c,d) Output from each MCMC chain, highlight
    that we are sampling from the same stationary distribution}
  \label{fig:aphiddata}
\end{figure*}

Next we consider a model of aphid population dynamics as proposed in
\cite{matis2007stochastic}. The system can be represented by the following
reactions:
 \begin{align}
   \centering
  \begin{array}{rrcl}
    \mathbf{R}_{1}: & N & \rightarrow & 2N + C \\
    \mathbf{R}_{2}: & N + C & \rightarrow & C
  \end{array}
\end{align}
and summarised in terms of
its stoichiometry matrix $S$, and hazard function
$h(\mathbf{X}_{t},\theta)$,
\begin{align}
  S  &= \left(
    \begin{array}{rr}
      1 & -1 \\
      1 & 0
    \end{array} \right), &
  h(\mathbf{X}_{t},\theta) &= (\lambda N_{t}, \mu N_{t} C_{t}),
\end{align}
$\mathbf{X}_{0} = (N_{0},C_{0})$, $\theta = (\lambda,\mu)$.

$N_{t}$ and $C_{t}$ are the numbers of aphids alive at time $t$ and the
cumulative number of aphids that have lived up until time $t$ respectively. In
the first reaction, when a new aphid is introduced into the system we get an
increase in both the current number of aphids and the cumulative count. When an
aphid is removed the number of aphids decreases by one but the cumulative count
remains the same. Aphid death is considered to be proportional not only to the
number of aphids currently in the system, but also to the cumulative count,
representing the idea that over time they are making their own environment less
habitable.

Given initial conditions $\mathbf{X}_{0} = (N_{0},C_{0})$ and a set rate of
parameters $\theta$, we can simulate from the model. However the difficulty in practice with the
model presented here is that we will never have observations of the cumulative
count. Observations then are noisy, discrete time measurements of just a single
species of the system.

\subsubsection{Aphid data}
\label{sec:aphiddata}

We now consider the data described in \cite{matis2008mechanistic} consisting of
cotton-aphid counts for twenty seven treatment-block combinations. The
treatments consisted of three nitrogen levels (blanket, variable and none), three
irrigation levels (low, medium and high) and three blocks. The sampling times of
the data are $t = 0, 1.14,$ $2.29,3.57,4.57$ weeks, or every seven to eight
days. We restrict our investigation to a single treatment combination, three
data sets with blanket nitrogen level and low irrigation. If we denote the block
by $i \in \{1,2,3\}$ then the data $\mathcal{D}_{i}$ is the number of aphids,
$N$, in block $i$ at each time $t$. The data are plotted in
figure~\ref{fig:aphiddata}(a).

We make the assumption that the counts are observed with error such that
\begin{equation}
  \label{eq:aphiddata}
  d_{t} \sim \operatorname{Pois}(x_{t}),
\end{equation}
and use a set of weakly informative priors on the rate parameters $\theta$
\begin{equation}
  \label{eq:aphidprior}
   \log(\theta_{i}) \sim \mathcal{U}(-8,8), \quad \quad i = 1,2.
 \end{equation}
 We place a prior of the form
 \begin{equation}
   \label{eq:aphhidcprior}
   C_{0} = N_{0} + g, \quad g \sim \operatorname{Geom}(0.03),
 \end{equation}
to reflect the fact that we are unable to measure $C_{0}$.

 We treat the three sets of observations as repeats of the same experiment. \red{Likelihood estimates are obtained by running a particle filter for each of the three data sets and taking the product of the individual estimates.} A
 full treatment of all twenty-seven data sets using a fixed effects model can be
 found in \cite{gillespie2010bayesian}. We consider the initial aphid counts to
 be the true values, as in \cite{gillespie2010bayesian}, on the basis that there
 should be no error in counting such small populations.

\red{\subsubsection{Results for the aphid growth model}}

We use the same criteria for the choice of $\epsilon_{t}$ for the ABC section of
the inference as with the Lotka--Volterra model in \ref{sec:resultslv}, namely
the 0.3 quantile of the distribution of distances. A sequence of five
distributions gives us 1000 samples from
$\pi(\theta|\rho(\mathcal{D},\mathcal{D}^{\ast}) < \epsilon_{4})$ which we use to initialise eight parallel chains. We record 20,000 samples from the exact
target posterior $\pi(\theta|\mathcal{D})$ after appropriate thinning.
Figure~\ref{fig:aphiddata}(c and d) shows the analysis of the MCMC chains. Again
we find that each chain is sampling from the same target and posterior densities
are very close from all eight chains. Figure~\ref{fig:aphiddata}(b) shows
posterior predictive quantiles given a sample from the pooled posterior
distribution. Model fit appears to be reasonable. \red{The results are consistent with those seen in \cite{gillespie2010bayesian} where they assume that observations are made without error and make use of an approximate simulation algorithm for realisations of the model.}
\subsection{Gene expression}
\label{ssec:gene}

\begin{figure*}[t]
  \centering
  \includegraphics[width = \textwidth]{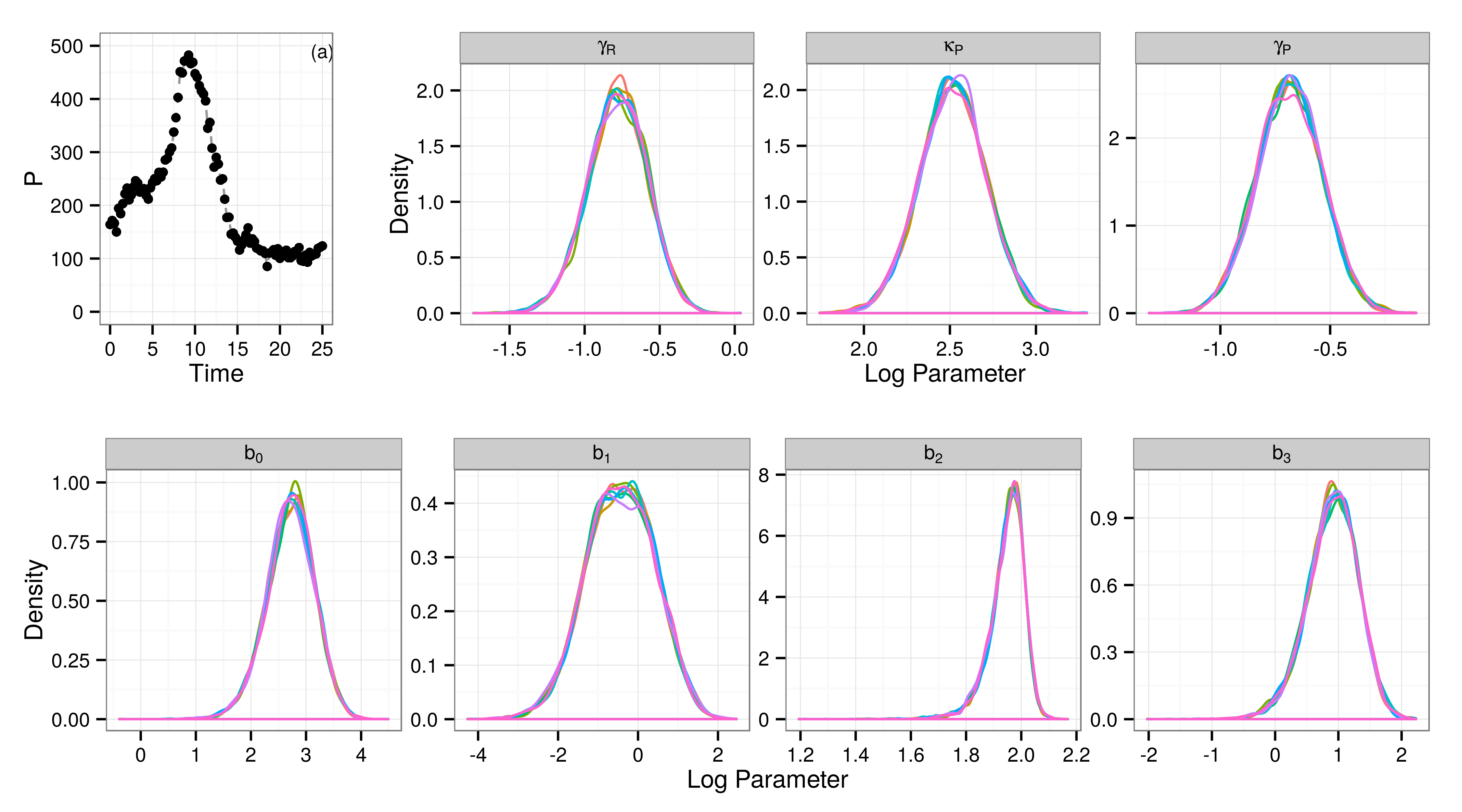}
  \caption{(a) is the noisy pseudo-data for the Protein levels in the model. The other plots show the individual densities from pMCMC chains after appropriate thinning having been initialised via an ABC run as described in section~\ref{ssec:tune}. The plots clearly show that each of the chains are in agreement with regard to sampling from the stationary distribution.}
  \label{fig:kom}
\end{figure*}

\begin{figure*}[t]
  \centering
  \includegraphics[width= \textwidth]{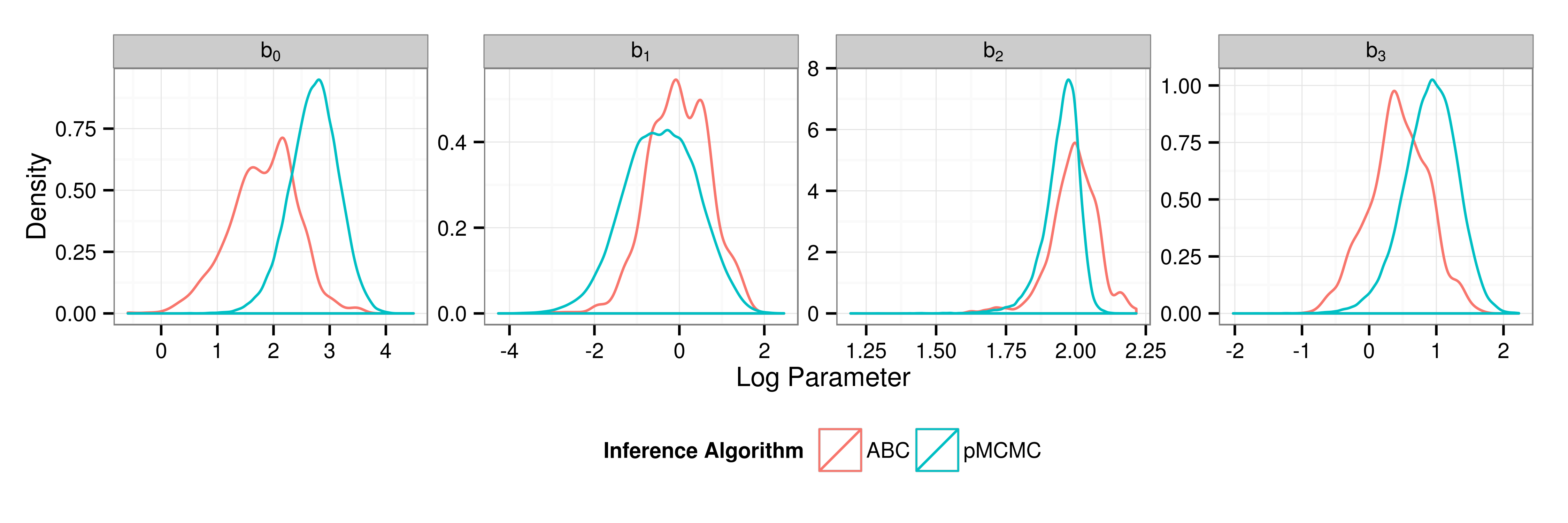}
  \caption{A comparison between the final sample using the ABC SMC algorithm and the pMCMC for the gene regulation model for the four parameters in the time dependent hazard. The plot shows that their is a distinct difference between the two posterior samples. Plots for the other three parameters show similar but are omitted here.}
  \label{fig:compare}
\end{figure*}

\red{
Finally, we consider a simple gene regulation model characterised by three species (DNA, mRNA, denoted R, and protein, P) and four reactions. The reactions represent transcription, mRNA degradation, translation and protein degradation respectively. The system has been analysed by \cite{komorowski2009bayesian} and \cite{golightly2014delayed} among others:
\begin{align}
 \begin{array}{rrcl}
    \mathbf{R}_{1}: & DNA & \rightarrow & DNA + R \\
    \mathbf{R}_{2}: & R & \rightarrow & \emptyset \\
    \mathbf{R}_{3}: & R & \rightarrow & R + P \\
    \mathbf{R}_{4}: & P & \rightarrow & \emptyset 
  \end{array}
\end{align}
with stoichiometry matrix $\mathbf{S}$, and hazard function $h(\mathbf{X}_{t},\theta)$
\begin{align}
  \mathbf{S} &= \left(
    \begin{array}{rrrr}
      1 & -1 & 0 & 0 \\
      0 & 0 & 1 & -1
    \end{array}
  \right), &
  h(\mathbf{X}_{t}, \theta) &= (\kappa_{R,t}, \gamma_{R}R_{t}, \kappa_{P}R_{t}, \gamma_{P}P_{t}) 
\end{align}
where $\mathbf{X}_{t} = (R_{t}, P_{t})$ and $\theta = (\gamma_{R}, \kappa_{P}, \gamma_{P}, b_{0}, b_{1}, b_{2}, b_{3})$ where we note that, as in \cite{komorowski2009bayesian}, we take $\kappa_{R,T}$ to be the time dependent transcription rate. Specifically,
\begin{align}
  \kappa_{R,t} &= b_{0}\exp(-b_{1}(t - b_{2})^{2}) + b_{3}
\end{align}
such that the transcription rate increases for $t < b_{2}$ and tends towards a baseline value, $b_{3}$, for $t > b_{2}$. As above the goal is inference on the unknown parameter vector, $\theta$. In keeping with inference in \cite{komorowski2009bayesian} we create a data poor scenario, 100 observations of synthetic data simulated given initial conditions $X_{0} = (10,150)$ and parameter values $(0.44,0.52,10,15,0.4,7,3)$ corrupted with measurement error, $Y_{t} \sim \mathcal{N}(X_{t},I\sigma^{2})$, $\sigma = 10$, with observations on the mRNA discarded. The data is shown in figure~\ref{fig:kom}(a).

We follow \cite{komorowski2009bayesian} by assuming the same prior distributions, including informative priors for the degradation rates to ensure identifiably. Specifically
\begin{align*}
  \gamma_{R} &\sim \Gamma(19.36,44) & \gamma_{P} &\sim \Gamma(27.04,52) \\
  \kappa_{P} &\sim \operatorname{Exp}(0.01) & b_{0} &\sim \operatorname{Exp}(0.01) \\
  b_{1} &\sim \operatorname{Exp}(1.0) &  b_{2} &\sim \operatorname{Exp}(0.1) \\
  b_{3} &\sim \operatorname{Exp}(0.01)
\end{align*}
where $\Gamma(a,b)$ is the gamma distribution with mean $a/b$ and $\operatorname{Exp}(a)$ is the Exponential distribution with mean 1/a.

For simplicity we assume that both the initial state, $X_{0} = (10,150)$, and the measurement error standard deviation, $\sigma = 10$, are known.

\subsubsection{Results for gene expression data}
\label{sssec:komresults}
We follow the same procedure as with the two examples above. Using a sequential ABC run to obtain a sample of 1000 parameters vectors distributed according to the approximate posterior. We then use eight random draws from the final ABC sample to initialise the parallel pMCMC chains with tuning parameters chosen as described in section~\ref{ssec:tune}. The posterior densities, a sample of 4000 from each chain having been subjected to appropriate thinning, are shown in figure~\ref{fig:kom}. It is clear that each of the chains is sampling from the same target giving us confidence in the resulting densities. The posteriors obtained are consistent with those in \cite{golightly2014delayed} and true parameter values are well identified. Figure~\ref{fig:compare} shows that the sample from the final iteration of the sequential ABC algorithm is markedly different from that in the pMCMC algorithm. There is a measurable improvement in using this type of scheme over using solely ABC in this way. We characterise the difference shown here as an improvement due to the fact that we know that pMCMC is asymptotically exact.}

\section{Discussion}
\label{sec:discussion}

\red{We have proposed an approach to inference for Markov processes that is asymptotically exact and
combines the relative strengths of ABC and pMCMC methodology to increase
computational efficiency through use of parallel hardware.} Through use of an
approximation to the posterior distribution of interest, obtained via a
sequential ABC algorithm which is easy to parallelise, we can set up a parallel
implementation of pMCMC which has numerous desirable properties. By enabling the
construction of independent parallel chains initialised close to the stationary
distribution, this enables fast convergence and sampling from an exact posterior
distribution that scales well with available computational resources. \red{Throughout our analyses we have made use of parallel computation, however we believe that the proposed approach will also be of interest in situations where parallel hardware is not available, as it still addresses the pMCMC initialisation and tuning problem.}
Algorithmic tuning parameters required for pMCMC, such as the variance of
Gaussian random walk proposals and numbers of particles for the particle filter
can be chosen without the need for additional pilot runs, as a consequence of
having a sample from an ABC posterior. In addition, independent parallel chains
allow verification of convergence and the computational saving in burn--in times
extends to repeat MCMC analyses.

We have demonstrated this approach by applying it to \red{three} stochastic kinetic
models. With the Lotka--Volterra predator prey system, a relatively simple model
in which both species can be observed, we highlighted clear issues with
practical implementation of a pseudo-marginal approach in a scenario in which
prior information on reaction rate parameters is poor. This issue can be
alleviated by first obtaining a sample from an approximation to the posterior,
then using it to guide an exact pMCMC scheme. The approach discussed performed
similarly well in the application to a set of real data for a model for aphid
growth dynamics in which one of the species in the system can never be observed,
where we again imposed weak prior conditions on the rate parameters governing
the system and had access to repeat data. \red{Finally we applied the scheme to a gene regulation model in which we had partially observed data and rate parameters were not all time homogeneous. The analyses of results show that we can verify that we are sampling from the same target distribution adding to our belief that we have converged to the true posterior of interest in each case.}

\section*{Acknowledgements}
The authors gratefully acknowledge the referees for their detailed and constructive comments. JO was supported by an EPSRC studentship.


\begin{thebibliography}{37}
\providecommand{\natexlab}[1]{#1}
\providecommand{\url}[1]{{#1}}
\providecommand{\urlprefix}{URL }
\expandafter\ifx\csname urlstyle\endcsname\relax
  \providecommand{\doi}[1]{DOI~\discretionary{}{}{}#1}\else
  \providecommand{\doi}{DOI~\discretionary{}{}{}\begingroup
  \urlstyle{rm}\Url}\fi
\providecommand{\eprint}[2][]{\url{#2}}

\bibitem[{Andrieu and Roberts(2009)}]{andrieu2009pseudo}
Andrieu C, Roberts GO (2009) The pseudo-marginal approach for efficient {Monte}
  {Carlo} computations. The Annals of Statistics 37(2):697--725

\bibitem[{Andrieu et~al(2010)Andrieu, Doucet, and
  Holenstein}]{andrieu2010particle}
Andrieu C, Doucet A, Holenstein R (2010) Particle {Markov} chain {Monte}
  {Carlo} methods. Journal of the Royal Statistical Society: Series B
  (Statistical Methodology) 72(3):269--342

\bibitem[{Beaumont et~al(2002)Beaumont, Zhang, and
  Balding}]{beaumont2002approximate}
Beaumont MA, Zhang W, Balding DJ (2002) Approximate {Bayesian} computation in
  population genetics. Genetics 162(4):2025--2035

\bibitem[{Beaumont et~al(2009)Beaumont, Cornuet, Marin, and
  Robert}]{beaumont2009adaptive}
Beaumont MA, Cornuet JM, Marin JM, Robert CP (2009) Adaptive approximate
  {Bayesian} computation. Biometrika 96(4):983--990

\bibitem[{Blum et~al(2013)Blum, Nunes, Prangle, Sisson
  et~al}]{blum2013comparative}
Blum MG, Nunes MA, Prangle D, Sisson SA, et~al (2013) A comparative review of
  dimension reduction methods in approximate bayesian computation. Statistical
  Science 28(2):189--208

\bibitem[{Boys et~al(2008)Boys, Wilkinson, and Kirkwood}]{boys2008bayesian}
Boys RJ, Wilkinson DJ, Kirkwood TB (2008) Bayesian inference for a discretely
  observed stochastic kinetic model. Statistics and Computing 18(2):125--135

\bibitem[{Del~Moral et~al(2006)Del~Moral, Doucet, and
  Jasra}]{del2006sequential}
Del~Moral P, Doucet A, Jasra A (2006) Sequential {Monte} {Carlo} samplers.
  Journal of the Royal Statistical Society: Series B (Statistical Methodology)
  68(3):411--436

\bibitem[{Doucet et~al(2001)Doucet, de~Freitas, and
  Gordon}]{doucet2001sequential}
Doucet A, de~Freitas N, Gordon N (2001) Sequential {Monte} {Carlo} methods in
  practice. Springer

\bibitem[{Doucet et~al(2012)Doucet, Pitt, and Kohn}]{doucet2012efficient}
Doucet A, Pitt M, Kohn R (2012) Efficient implementation of {Markov} chain
  {Monte} {Carlo} when using an unbiased likelihood estimator. arXiv:12101871

\bibitem[{Drovandi and Pettitt(2011)}]{drovandi2011estimation}
Drovandi CC, Pettitt AN (2011) Estimation of parameters for macroparasite
  population evolution using approximate {Bayesian} computation. Biometrics
  67(1):225--233

\bibitem[{Fearnhead and Prangle(2012)}]{fearnhead2012constructing}
Fearnhead P, Prangle D (2012) Constructing summary statistics for approximate
  {Bayesian} computation: semi-automatic approximate {Bayesian} computation.
  Journal of the Royal Statistical Society: Series B (Statistical Methodology)
  74(3):419--474

\bibitem[{Filippi et~al(2013)Filippi, Barnes, Cornebise, and
  Stumpf}]{filippi2013optimality}
Filippi S, Barnes CP, Cornebise J, Stumpf MP (2013) On optimality of kernels
  for approximate bayesian computation using sequential monte carlo.
  Statistical applications in genetics and molecular biology 12(1):87--107

\bibitem[{Gillespie(2009)}]{Gillespie2009a}
Gillespie CS (2009) {Moment-closure approximations for mass-action models}. IET
  Systems Biology 3(1):52--8, \doi{10.1049/iet-syb:20070031}

\bibitem[{Gillespie and Golightly(2010)}]{gillespie2010bayesian}
Gillespie CS, Golightly A (2010) Bayesian inference for generalized stochastic
  population growth models with application to aphids. Journal of the Royal
  Statistical Society: Series C (Applied Statistics) 59(2):341--357

\bibitem[{Gillespie(1977)}]{gillespie1977exact}
Gillespie DT (1977) Exact stochastic simulation of coupled chemical reactions.
  The Journal of Physical Chemistry 81(25):2340--2361

\bibitem[{Gillespie(2000)}]{gillespie2000chemical}
Gillespie DT (2000) The chemical {Langevin} equation. The Journal of Chemical
  Physics 113(1):297--306

\bibitem[{Golightly and Wilkinson(2011)}]{golightly2011bayesian}
Golightly A, Wilkinson DJ (2011) Bayesian parameter inference for stochastic
  biochemical network models using particle {Markov} chain {Monte} {Carlo}.
  Interface Focus 1(6):807--820

\bibitem[{Golightly et~al(2014)Golightly, Henderson, and
  Sherlock}]{golightly2014delayed}
Golightly A, Henderson D, Sherlock C (2014) Delayed acceptance particle mcmc
  for exact inference in stochastic kinetic models. Statistics and Computing pp
  1--17, \doi{10.1007/s11222-014-9469-x},
  \urlprefix\url{http://dx.doi.org/10.1007/s11222-014-9469-x}

\bibitem[{Kitano(2002)}]{kitano2002computational}
Kitano H (2002) Computational systems biology. Nature 420(6912):206--210

\bibitem[{Komorowski et~al(2009)Komorowski, Finkenst{\"a}dt, Harper, and
  Rand}]{komorowski2009bayesian}
Komorowski M, Finkenst{\"a}dt B, Harper CV, Rand DA (2009) Bayesian inference
  of biochemical kinetic parameters using the linear noise approximation. BMC
  Bioinformatics 10(1):343

\bibitem[{Lotka(1925)}]{lotka1925elements}
Lotka AJ (1925) Elements of physical biology. Williams \& Wilkins Baltimore

\bibitem[{Marjoram et~al(2003)Marjoram, Molitor, Plagnol, and
  Tavar{\'e}}]{marjoram2003markov}
Marjoram P, Molitor J, Plagnol V, Tavar{\'e} S (2003) Markov chain {Monte}
  {Carlo} without likelihoods. Proceedings of the National Academy of Sciences
  100(26):15,324--15,328

\bibitem[{Matis et~al(2007)Matis, Kiffe, Matis, and
  Stevenson}]{matis2007stochastic}
Matis JH, Kiffe TR, Matis TI, Stevenson DE (2007) Stochastic modeling of aphid
  population growth with nonlinear, power-law dynamics. Mathematical
  Biosciences 208(2):469--494

\bibitem[{Matis et~al(2008)Matis, Parajulee, Matis, and
  Shrestha}]{matis2008mechanistic}
Matis TI, Parajulee MN, Matis JH, Shrestha RB (2008) A mechanistic model based
  analysis of cotton aphid population dynamics data. Agricultural and Forest
  Entomology 10(4):355--362

\bibitem[{Pitt et~al(2012)Pitt, Silva, Giordani, and Kohn}]{pitt2012some}
Pitt MK, Silva RdS, Giordani P, Kohn R (2012) On some properties of {Markov}
  chain {Monte} {Carlo} simulation methods based on the particle filter.
  Journal of Econometrics 171(2):134--151

\bibitem[{Pritchard et~al(1999)Pritchard, Seielstad, Perez-Lezaun, and
  Feldman}]{pritchard1999population}
Pritchard JK, Seielstad MT, Perez-Lezaun A, Feldman MW (1999) Population growth
  of human y chromosomes: a study of y chromosome microsatellites. Molecular
  Biology and Evolution 16(12):1791--1798

\bibitem[{Roberts and Rosenthal(2001)}]{roberts2001optimal}
Roberts GO, Rosenthal JS (2001) Optimal scaling for various
  {Metropolis}-{Hastings} algorithms. Statistical science 16(4):351--367

\bibitem[{Roberts et~al(1997)Roberts, Gelman, and Gilks}]{roberts1997weak}
Roberts GO, Gelman A, Gilks WR (1997) Weak convergence and optimal scaling of
  random walk metropolis algorithms. The Annals of Applied Probability
  7(1):110--120

\bibitem[{Salis and Kaznessis(2005)}]{Salis2005}
Salis H, Kaznessis Y (2005) Accurate hybrid stochastic simulation of a system
  of coupled chemical or biochemical reactions. Journal of Chemical Physics
  122:054,103

\bibitem[{Sherlock et~al(2013)Sherlock, Thiery, Roberts, and
  Rosenthal}]{sherlock2013efficiency}
Sherlock C, Thiery AH, Roberts GO, Rosenthal JS (2013) On the efficiency of
  pseudo-marginal random walk {Metropolis} algorithms. arXiv:13097209

\bibitem[{Silk et~al(2013)Silk, Filippi, and Stumpf}]{silk2013optimizing}
Silk D, Filippi S, Stumpf MP (2013) Optimizing threshold-schedules for
  sequential approximate {Bayesian} computation: applications to molecular
  systems. Statistical applications in genetics and molecular biology
  12(5):603--618

\bibitem[{Sisson et~al(2007)Sisson, Fan, and Tanaka}]{sisson2007sequential}
Sisson S, Fan Y, Tanaka MM (2007) Sequential {Monte} {Carlo} without
  likelihoods. Proceedings of the National Academy of Sciences
  104(6):1760--1765

\bibitem[{Tavare et~al(1997)Tavare, Balding, Griffiths, and
  Donnelly}]{tavare1997inferring}
Tavare S, Balding DJ, Griffiths R, Donnelly P (1997) Inferring coalescence
  times from dna sequence data. Genetics 145(2):505--518

\bibitem[{Toni et~al(2009)Toni, Welch, Strelkowa, Ipsen, and
  Stumpf}]{toni2009approximate}
Toni T, Welch D, Strelkowa N, Ipsen A, Stumpf MP (2009) Approximate {Bayesian}
  computation scheme for parameter inference and model selection in dynamical
  systems. Journal of the Royal Society Interface 6(31):187--202

\bibitem[{Volterra(1926)}]{volterra1926fluctuations}
Volterra V (1926) Fluctuations in the abundance of a species considered
  mathematically. Nature 118:558--560

\bibitem[{Wilkinson(2006)}]{wilkinson2006parallel}
Wilkinson DJ (2006) Parallel {Bayesian} computation, Statistics Textbooks and
  Monographs, vol 184. MARCEL DEKKER AG

\bibitem[{Wilkinson(2011)}]{wilkinson2011stochastic}
Wilkinson DJ (2011) Stochastic modelling for systems biology, Chapman \&
  Hall/CRC mathematical biology and medicine series, vol~44, 2nd edn. CRC press

\end{thebibliography}
\end{document}